\documentstyle[12pt]{article}
\def\lromn#1{\uppercase\expandafter{\romannumeral#1}}
\def\romn#1{\romannumeral#1}

\def\blist{\begin{list}{\setlength{\rightmargin}{\leftmargin}}}
\def\elist{\end{list}}
\begin{document}
\begin{flushright}
TU/96/500
\end{flushright}

\vspace{12pt}

\begin{center}
\begin{Large}

\renewcommand{\thefootnote}{\fnsymbol{footnote}}
\bf{
Baryogenesis and thermal history after inflation
\footnote[1]
{
Lecture delivered at KOSEF-JSPS Winter School, held at Seoul, Korea,
February 21-28, 1996. To appear in the Proceedings to be published
in Journal of Korean Physical Society.
}
}
\end{Large}

\vspace{1cm}

\begin{large}
M. Yoshimura \\
Department of Physics, Tohoku University\\
Sendai 980-77 Japan\\
\end{large}

\vspace{5cm}

{\bf ABSTRACT}
\end{center}

The basic idea of baryogenesis is lectured to introduce non-experts
to this subject.
Some recent topics, necessarily subjective in view of short time
limitation, are also presented to show how the initial condition
for baryogenesis is realized in the new framework of inflation.

\newpage

\section{Introduction}
Let us first recall the modern view of cosmology; big bang theory.
According to general relativity the spacetime evolution is determined
via the Einstein equation 
by the matter content of universe, which differs from epoch to epoch
depending on what kind of energy dominates the energy density
of the universe at that time.
There are three important epochs characterized by different 
pressure-energy relation;
(1) vacuum energy dominance with
\( \:
p = -\,\rho 
\: \),
(2) massless particle dominance with
\( \:
p = \frac{1}{3}\rho 
\: \),
(3) nonrelativistic particle dominance with
\( \:
p \ll \rho \,.
\: \)
With homogeneity and isotropy of 3-space the Einstein equation is
much simplified with the Robertson-Walker metric:
\( \:
ds^{2} = dt^{2} - a^{2}(t)\,\vec{x}^{2} \,.
\: \)
The behavior of the scale factor $a(t)$ then follows;
\begin{eqnarray}
&&
(1) \hspace{0.5cm} 
a \:\propto  \: e^{Ht} \,, \hspace{0.3cm} H = \sqrt{8\pi G V/3}
 \,, \\
&&
(2) \hspace{0.5cm} 
a \:\propto  \: t^{1/2}
\,, \\
&&
(3) \hspace{0.5cm} 
a \:\propto  \: t^{2/3} \,, 
\end{eqnarray}
in their respective energy ranges.
The first stage is the inflationary epoch where a constant vacuum
energy $V$ gives the exponential growth of the scale factor, which
is believed to solve the horizon and the flatness problems in
the old big bang model \cite{inflation-rev}.

It would be of some interest to view the Hubble parameter defined by
\( \:
H = \dot{a}/a
\: \)
as a function of $a$ or a typical energy scale $T\propto 1/a$ of
dominant massless particles in the universe,
instead of $a(t)$ as a function of time.
In this picture the first inflationary stage is longest, lasting
for the scale change $>10^{30}$.
Subsequent radiation and matter dominant epochs have energy changes of
order, $10^{26}$ or $10^{3}$.
The standard model of microphysics only probes the temperature range
of
\( \:
10^{2} \,{\rm GeV} \:- \: 10^{-13} \,{\rm GeV} \,, 
\: \)
of much smaller variation than in the inflationary epoch.
Of great importance in subsequent cosmic
evolution and of intensive current interest
is the transient stage from inflation to radiation dominance;
the epoch of reheating after inflation, which I shall touch upon later
in this lecture.

For a long time the success of the big bang model rested with
three cornerstones;\\
\hspace*{0.5cm} 
(1) \hspace{0.3cm} Hubble expansion, \\
\hspace*{0.5cm} 
(2) \hspace{0.3cm} Planck distribution of relic photons,\\
\hspace*{0.5cm} 
(3) \hspace{0.3cm} light element abundance such as $^{4}$He, D, $^{7}$Li.\\
In the last decade there have been substantial improvements 
of this classic achievement both in observation and in theory.
Perhaps the most significant is COBE measurement of the spectrum
shape and large scale fluctuation of the relic microwave
temperature: absolute motion of earth relative to the universal
cosmic expansion has been observed with
\( \:
\delta T/T \approx 10^{-3} \,, 
\: \)
and after subtraction of this dipole component there exists a large
scale fluctuation \cite{3kfluc-obs} 
of order $10^{-6}$, whose precise feature yet to
be clarified is very important in the theory of structure formation.
Nucleosynthesis \cite{nucleosyn-rev} 
has become a precise science developed to a fine
detail, so much refined that one can even discuss dependence on
the number of neutrino species and the neutron lifetime.
But presumably the most important in nucleosynthesis
is a consistent and reliable determination of the key cosmological
parameter, the baryon to photon ratio $n_{B}/n_{\gamma }$,
of order $10^{-10}$.

In order to probe microphysical processes that occur at each instant
of cosmological time, it is necessary to compare the two time scales;
physical time scale during which a relevant process takes place, and
the Hubble time scale of order $1/H$. If the time scale of a physical
process is much shorter than the Hubble time, then this physical
process will occur frequently, and in the limit of large rate 
equilibrium is reached.

In a weak coupling theory such as supersymmetric grand unified theories
beyond the TeV scale, the proper framework to discuss the interaction rate
is the Boltzmann equation for one particle distribution function
taking into account  the cosmological expansion.
When the Boltzmann equation is integrated with respect to the momentum
phase space, an equation for the number density follows;
\begin{eqnarray}
(\,\frac{d}{dt} + 3\frac{\dot{a}}{a}\,)\,n_{i} = 
\int\,\frac{d^{3}p_{i}}{(2\pi )^{3}}\,\frac{\Gamma _{i}[f]}{E_{i}}
\,, 
\end{eqnarray}
where $\Gamma _{i}[f]$ is the invariant interaction kernel that depends on
all distribution functions $f_{j}$ participating in the process.
As an example let me give the form of the kernel when only
two-body reactions are involved,
\begin{eqnarray}
\Gamma _{i}[f]&=& \frac{1}{2} \sum_{j\,, k\,, l}
\int\,d{\cal P}_{j}d{\cal P}_{k}d{\cal P}_{l}\,[\, 
(1 \pm f_{i})(1 \pm f_{j})f_{k}f_{l}\,\gamma _{kl\rightarrow ij} 
\nonumber \\
&& \hspace*{2cm}
-\,(1 \pm f_{k})(1 \pm f_{l})f_{i}f_{j}\,\gamma _{ij\rightarrow kl}\,]
\,, 
\end{eqnarray}
where $\gamma _{ij\rightarrow kl}$ is the invariant rate for the process
$ij \rightarrow kl$ essentially given by the differential cross section times
the initial relative velocity, and $\pm $ corresponds to either
stimulated emission for bosons or Pauli blocking for fermions.
$d{\cal P} = \frac{d^{3}p}{E(2\pi )^{3}}$
is the invariant phase space element.

In general it is difficult to solve non-linear integro-differential
equations of the Boltzmann type, but it is often possible to estimate
reaction rate in the right hand side by some means.
In certain situations the rate may roughly vary linearly with
the number density itself $n_{i}$;
\begin{equation}
(\,\frac{d}{dt} + 3H\,)\,n_{i} = 
-\,\overline{\gamma_{r} }\,n_{i} \,,
\end{equation}
with $H = \dot{a}/a$.
For instance, if the relevant particle freely decays having no interaction 
with medium particles and no process of particle production operates,
$\overline{\gamma_{r} }$ is simply the free decay rate.

A particular reaction is of cosmological relevance 
only if the reaction rate given by
the right hand side of the Boltzmann equation is much larger than
the Hubble rate $3Hn_{i}$. What typically happens is that
the process continues at early times and after some late time it
is frozen.
In cosmology it is thus important to estimate decoupling epoch of a
particular reaction.
The crucial quantity in determining relevance of the reaction is
the ratio of the average rate to the Hubble rate 
$\overline{\gamma_{r} }/H$.
If this ratio is very large, the reaction occurs frequently.

As an important practical example let us consider the four Fermi type
interaction governed by a coupling $\alpha /M^{2}$.
This could be the weak interaction involving neutrinos such as
\( \:
\nu \bar{\nu } \:\rightarrow  \: e^{+}e^{-}
\: \)
or baryon number violating reaction, 
\( \:
qq \:\rightarrow  \: \bar{q}\bar{l} \,,
\: \)
if one uses as the relevant mass  $M$ the electroweak or the GUT 
mass scale.
Without any detailed discussion I would give you a suggestive form
of the average rate,
\begin{eqnarray}
\overline{\gamma_{r} } \approx \frac{\alpha ^{2}T^{2}}{(T^{2} + M^{2})^{2}}
\cdot T^{3} \,.
\end{eqnarray}
Here $T$ is a typical energy of participating particles and of order
the temperature if all these particles are in thermal and chemical
equilibrium, and the last factor $T^{3}$ represents the number density
of colliding particles.
The factor in front of this number density is the averaged reaction
rate for the elementary process, which
grows with energy $T$ as $\alpha ^{2}T^{2}/M^{4}$ for $T \ll M$, but
is finally saturated and decreases as $\alpha ^{2}/T^{2}$ in
the high energy limit.
On the other hand, the Hubble rate is of order $\sqrt{N}T^{2}/m_{{\rm pl}}$
with 
\( \:
m_{{\rm pl}} = 1/\sqrt{G} \sim 10^{19}\,
\: \)
GeV the Planck mass.
$N$ is the number of massless particle species in equilibrium.
It is then easy to see that the ratio $\overline{\gamma _{r}}/H$ has a maximum
at around $T = M$ and the maximum ratio is of order 
\begin{equation}
\alpha ^{2}m_{{\rm pl}}/(\sqrt{N}M) \,.
\end{equation}
The condition of the large rate for equilibrium is thus
\( \:
M \ll \alpha ^{2}m_{{\rm pl}}/\sqrt{N} \,.
\: \)
Since the right hand side contains the Planck mass, this condition
is easily obeyed for the weak processes.
This is an important observation to understand why the element abundance
cooked up in the early universe is insensitive to the initial condition:
one can justify that the initial neutron-proton ratio is
given by the value in thermal and chemical equilibrium.

As another example let us consider
how much of baryons is left over if the universe is symmetrical 
with respect to baryons and anti-baryons.
This is a problem relevant to baryogenesis, the main theme of this lecture.
As the final element carrying the negative baryon number, 
anti-nucleons gradually pair-annihilate via many pion production;
\( \:
N\overline{N} \rightarrow  {\rm many}\;\pi{\rm '} \,{\rm s}\,,
\: \)
but after the decoupling of this annihilation process the leftover
abundance $n_{N}/n_{\gamma } = n_{\overline{N}}/n_{\gamma }$ is frozen.
The annihilation process has a cross section roughly of order
\( \:
\langle \sigma v \rangle \approx 1/m_{\pi }^{2} 
\: \)
at low energies,
independent of the energy participating in the process.
On the other hand, the thermal number density is suppressed by
the Boltzmann factor when nucleons and anti-nucleons become nonrelativistic
as actually is the case;
\begin{equation}
n_{N} = n_{\overline{N}} \sim 
4\,(\frac{m_{N}T}{2\pi })^{3/2}\,e^{-\,m_{N}/T}\,.
\end{equation}
From the decoupling condition,
\( \:
\langle \sigma v \rangle\cdot n_{N} = H \,, 
\: \)
one then finds for the decoupling temperature $T_{d}$ and the leftover
abundance that
\begin{eqnarray}
&&
\frac{T_{d}}{m_{N}} \approx  [\,\ln \frac{m_{N}m_{{\rm pl}}
\langle \sigma v \rangle}{60\sqrt{N}}\,]^{-1} \sim \frac{1}{45}
 \,, \\
&&
\frac{n_{N}}{n_{\gamma }} \approx  \frac{40\sqrt{N}}{m_{N}m_{{\rm pl}}
\langle \sigma v \rangle} \sim 10^{-18} \,.
\end{eqnarray}
The final abundance $n_{N}/n_{\gamma }$
is too small compared with the observed value $\approx  10^{-10}$, and
this argument essentially rules out the symmetric cosmology.

In the asymmetric universe with respect to baryons and anti-baryons
the present number ratio $n_{B}/n_{\gamma }$ is the measure of
the imbalance between baryons and anti-baryons prior to the
annihilation process:
\begin{equation}
\left( \frac{n_{B}}{n_{\gamma }}\right)_{{\rm present}} \approx 
\left(\frac{B - \overline{B}}{B + \overline{B}} \right)
_{{\rm before \; annihilation}} \,, 
\end{equation}
since in thermal and chemical equilibrium 
baryons, anti-baryons, and photons are all
in roughly equal abundance,
\( \:
B \approx \overline{B} \approx \gamma \,, 
\: \) 
ignoring the mass threshold and the statistical factor.
Our task is thus to explain this quantity from microphysics.

\section{Condition for baryogenesis}
Although necessary ingredients for the baryogenesis were written
a long time ago\cite{sakharov}, 
intricacy of this condition has been spelled out
much later \cite{my78},\cite{ttez},\cite{wein79},\cite{dimosuss}.
(For reviews, see \cite{basym-review-mine},\cite{basym-review-kolbt}.)
Their fascinating features  still invite many interesting scenarios.
The three obvious conditions are \\
\hspace*{0.5cm} (1) $\;$ baryon number non-conservation,\\
\hspace*{0.5cm} (2) $\;$
violation of dicrete ${\cal C}$ and ${\cal CP}$ symmetry,\\
\hspace*{0.5cm} (3) $\;$ departure from thermal equilibrium.

The baryon number is not a sacred symmetry in modern gauge theories,
and indeed violated explicitly by heavy $X$ boson mediated processes 
in grand unified theories \cite{gut-rev-langacker}.
Moreover, even in the standard $SU(3)\times SU(2) \times U(1)$
gauge theory the baryon number is violated 
by instanton effects at zero temperature\cite{thooft-76}, 
which is however unobservably suppressed.
It has however been recognized that 
sphaleron-mediated processes \cite{krs-ew-bgen}
may enhance baryon non-conservation in the standard theory 
at high enough temperatures, higher than TeV scale, although $B - L$
is exactly conserved here.
The sphaleron \cite{sphaleron} is a finite energy field configuration 
which bridges between different nonperturbative vacuum configurations 
ordinarily suppressed by tunneling probability at zero temperature.
The effect may be enhanced at a high temperature $T$ by the factor
$e^{-F/T}$, with $F$ the free energy of this unstable sphaleron.
The sphaleron effect can readily wash out
the baryon asymmetry generated prior to this epoch if initially
$B - L = 0$.
It should however be noted that it is not a trivial matter
to create from a symmetric state the asymmetry at the electroweak scale.
This is related to that
the last two conditions are much more subtle, as I shall explain shortly.

Is the standard model capable of explaining the baryon asymmetry?
First of all, the first two conditions are met as a matter of principle
even in the standard model, 
although strength of $B$ violation is to be detailed. 
The last condition of departure from
equilibrium is in general difficult to meet.
Most scenarios \cite{ew-bgeneration-review} 
use the first order phase transition of the electroweak
gauge symmetry breaking as a means of setting up non-equilibrium environment.
This requires a Higgs boson mass $< 45\,$GeV in the standard model of
one Higgs doublet, which 
already seems to be ruled out by LEP experiments. 
But this bound is based on a one-loop computation,
so there might be some loophole in the argument not contemplated so far,
and it may be worthwhile to explore
the possibility of baryogenesis, ignoring the Higgs mass bound.

Farrar and Shaposhnikov \cite{farrar-shapo} 
recently proposed an interesting scenario
that employs quark scattering off the electoweak bubble created at
the first order phase transition.
As an idealization one may assume that baryon number is strongly
violated in the unbroken phase, while it is conserved in the broken phase.
Quarks are scattered off within some length near the bubble wall and
there may be many scattering amplitudes that interfere, giving non-trivial
${\cal CP}$ violating effect as required.
Unfortunately the original calculation neglected the important effect
of coherence in the cosmic plasma \cite{basym-standard-th}. 
Coherence is crucial to generate
the baryon asymmetry by this mechanism and is only maintained by
the coherence length which is severely limited in the cosmic plasma.
The resultant asymmetry is too small to
yield the one needed by observation, typically
\( \:
< 10^{-20} 
\: \) \cite{basym-standard-th}.
Thus at the moment there is no viable model of baryogenesis that explains
the observed ratio using only the source of ${\cal CP}$ violation identical
to the CKM matrix.

According to our present wisdom the standard model must be extended 
in order to explain the baryon asymmetry.
There are two attitudes in extending the established physics with this
respect: one is to use the new physics of unification such as grand
unified theories or its supersymmetric extention that have
the natural source of all ingredients for baryogenesis, and the other 
\cite{ew-bgeneration-review} is
to minimize the extention utilizing the strong electroweak baryon
violation maximally.
The way I would like to characterize these two directions is that
the first direction needs a big jump beyond the standard theory,
while its physical mechanism of baryogenesis is straightforward and
readily understandable.
On the other hand, the second extention needs a small step beyond
what we already know, such as a slight modification of the Higgs system, 
but physics involved is fairly complicated, 
and there seems no model everyone would like to consider seriously.
With this situation in mind I would like to mainly discuss essential features
of baryogensis based on GUT.
I shall also discuss obstacle against the GUT scenario and 
how to evade it.

Although not discussed below, there is another interesting scenario
of baryogenesis that is becoming popular recently; the Affleck-Dine
mechanism \cite{affleck-dine}.
This mechanism uses a feature of supersymmetric models;
existence of many flat field directions in the scalar field potential.
These directions may include field condensate that carrys the baryon
number or the lepton number, both of which can be used when combined with
the strong electroweak baryon nonconservation at finite temperatures.
The resulting baryon to photon ratio tends to be large compared to
the observed one, and one usually has to consider some process of
dilution at the same time.
A version of this mechanism is discussed in the lecture by L. Randall at
this school.

\section{GUT baryogenesis}
Baryon nonconserving processes in grand unified theories
are mediated by gauge or Higgs bosons generically called $X$ bosons.
The simplest and workable scenario of GUT baryogenesis employs $X$ boson
decay, which has two types of decay modes, the two quark decay mode
$qq$, and the leptoquark mode $\bar{q}\bar{l}$.
Coexistence of the two modes with different baryon numbers is
a manifestation of baryon nonconservation.

Two simple examples of grand unified theories are $SU(5)$ and $SO(10)$
models \cite{gut-rev-langacker}. 
These models have unified multiplet structure of one family of
quarks and leptons of the form,
\begin{eqnarray}
SU(5) ; \hspace{0.3cm}
&&
\underline{5} + \underline{10} \,, \nonumber \\
&& \hspace*{-1cm}
\left( \begin{array}{c}
d_{R}  \\
d_{G} \\
d_{B} \\
e^{+} \\
\bar{\nu }
\end{array}
\right)_{R}
+
\left( \begin{array}{ccccc}
0 & \overline{u_{B}} & - \overline{u_{G}} & u_{R} & d_{R} \\
 & 0 & \overline{u_{R}} & u_{G} & d_{G} \\
 & & 0 & u_{B} & d_{B} \\
 & & & 0 &e^{+}  \\
&&& &0 
\end{array}
\right)_{L}
\,, \\
SO(10) ; \hspace{0.3cm}
&& 
\underline{16} \,, \nonumber \\
&& \hspace*{-1cm}
\left( \begin{array}{cccc}
u_{R} & u_{G} & u_{B} & \nu  \\
d_{R} & d_{G} & d_{B} & e
\end{array}
\right)_{L} +
\left( \begin{array}{cccc}
\overline{u_{R}} & \overline{u_{G}} & \overline{u_{B}} & \overline{N} \\
\overline{d_{R}} & \overline{d_{G}} & \overline{d_{B}} & \overline{e} 
\end{array}
\right)_{L} \,.
\end{eqnarray}
For instance, in the $SU(5)$ model 
$X$ gauge bosons induce transitions from a quark to a lepton in the 
representation $\underline{5}$, while in $\underline{10}$ they cause
transitions both from a quark to a lepton and from an anti-quark to a quark.

Recent precise measurements of coupling constants at LEP suggest 
\cite{susy-unification} 
that supersymmetric extention of the $SU(5)$ model gives a consistent
picture of coupling unification using the renormalization group.
The $SO(10)$ model, on the other hand, is interesting if a finite neutrino
mass hinted by the solar neutrino experiments is real.
Despite of the lack of evidence for proton decay grand unified theories
thus deserve serious consideration.

In discussing the GUT baryogenesis,
I shall dismiss complication due to many decay channels and concentrate
on the two decay modes for simplicity.
Let us denote decay rates of the two modes by
\( \:
\gamma _{q} \,, \;\gamma _{l}
\: \)
for $X$ and 
\( \:
\bar{\gamma }_{q} \,, \; \bar{\gamma }_{l}
\: \)
for its anti-particle $\bar{X}$.
When a pair of $X$ and $\bar{X}$ decays, a finite baryon number may
be created with a rate,
\begin{eqnarray}
\Delta B = \frac{2}{3}\gamma _{q} - \frac{1}{3}\gamma _{l}
- \frac{2}{3}\bar{\gamma }_{q} + \frac{1}{3}\bar{\gamma }_{l} 
= \gamma _{q} - \bar{\gamma }_{q} \,.
\end{eqnarray}
This quantity was simplified by using the requirement of ${\cal CPT}$
theorem, which states the equality of particle and anti-particle total
decay rate; 
\begin{equation}
\gamma _{q} + \gamma _{l} = \bar{\gamma }_{q} + \bar{\gamma }_{l} \,.
\end{equation}
Clearly ${\cal CP}$ violation is called for 
$\gamma _{q} \neq \bar{\gamma }_{q}$.

Before discussing perturbative calculation of this asymmetry,
I shall explain the non-equilibrium condition for the GUT baryogenesis.
Departure from equilibrium demands in this case that the inverse
decay process, 
\( \:
qq \rightarrow X \,, \;
\bar{q}\bar{l} \rightarrow X 
\: \)
etc. is essentially frozen by threshold effect.
This is realized if at the decay time the temperature is too low 
to create heavy $X$'s:
\( \:
T < m_{X}
\: \)
when the Hubble 
\( \:
H(T) = \gamma _{X}
\: \).
This leads to
\( \:
O[\sqrt{m_{{\rm pl}}\,\gamma _{X}\,}] < m_{X}
\: \).
Since the total decay rate $\gamma _{X} \sim \alpha\, m_{X}$,
it gives a constraint on the $X$ mass,
\begin{eqnarray}
m_{X} > O[\alpha\, m_{{\rm pl}}] \,.
\end{eqnarray}
A more precise estimate yields the $X$ mass bound of order 
\( \:
10^{15} - 10^{16} \,
\: \)
GeV, close to the GUT scale.
This constraint may be obeyed without much difficulty by a Higgs
$X$ boson, although with some difficulty by the gauge $X$ boson.

Calculation of the magnitude of the baryon asymmetry due to the $X$ boson
decay involves interesting interference effect.
Suppose that one computes the baryon production rate $\Delta B$ in
perturbation theory;
\begin{eqnarray}
\hspace*{-0.5cm}
\Delta B &=&\! \sum_{{\rm phase \,space}}\! 
|g_{1}f_{1} + g_{2}f_{2} + \cdots |^{2} - 
|g_{1}^{*}f_{1} + g_{2}^{*}f_{2} + \cdots |^{2} 
 \nonumber \\
&=&
-\,4\,\Im (g_{1}g_{2}^{*})\cdot\! \sum_{{\rm phase \,space}}\! 
\Im (f_{1}f_{2}^{*}) + \cdots 
\,.
\end{eqnarray}
Each decay amplitude $g_{i}f_{i}$ corresponds to a Feynman diagram and
$g_{i}$ lumps together all coupling factors leaving the dynamical
part of the amplitude to $f_{i}$. 
The rate for $\bar{X}$ thus has the complex-conjugated
coupling $g_{i}^{*}$, as required by ${\cal CPT}$ or hermiticity of
the Hamiltonian.
The final expression of $\Delta B$ clearly indicates need for a dynamical
phase 
\( \:
\Im (f_{1}f_{2}^{*}) \neq 0
\: \)
besides the ${\cal CP}$ violation,
\( \:
\Im (g_{1}g_{2}^{*}) \neq 0 \,.
\: \)
The dynamical phase may arise as a rescattering phase in ordinary
two-body processes, and one may regard the phase above as a generalization
of the rescattering phase.
The constraint for a non-trivial dynamical 
phase is that intermediate states must have
a threshold below the parent particle mass $m_{X}$.
This observation also hints an efficient way to compute the asymmetry 
as a discontinuity
according to the Landau-Cutkovsky rule: one can put intermediate particles
on the mass shell which makes actual computation of the asymmetry
much easier.

The baryon to photon ratio, or more precisely the baryon to 
entropy ratio $n_{B}/s$ may be derived from the baryon
production rate $\Delta B$ as follows.
Define first the baryon number in the comoving volume,
${\cal N}_{B} = a^{3}\,n_{B}$, together with a similar $X$ density,
${\cal N}_{X} = a^{3}\,n_{X}$. These vary as
\begin{equation}
\dot{{\cal N}}_{B} = \Delta B\cdot {\cal N}_{X} \,, \hspace{0.3cm}
\dot{{\cal N}}_{X} = -\,\gamma _{{\rm tot}}\cdot {\cal N}_{X} \,, 
\end{equation}
with $\gamma _{{\rm tot}} = \gamma _{q} + \gamma _{l}$ the total $X$
decay rate.
This is integrated with the condition that initially no baryon number
exists and finally no $X$ boson exists:
\begin{equation}
({\cal N}_{B})_{f} = \epsilon \,({\cal N}_{X})_{i} \,, \hspace{0.3cm}
\epsilon \equiv \frac{\Delta B}{\gamma _{{\rm tot}}} \,.
\end{equation}
Since this reaction and subsequent evolution proceeds 
almost adiabatically, the entropy per comoving volume is 
approximately conserved;
\( \:
a^{3}s =
\: \)
constant. This relation can be used to eliminate the volume factor $a^{3}$
in favor of the entropy density $s$;
\begin{equation}
(\frac{n_{B}}{s})_{f} = \epsilon\, (\frac{n_{X}}{s})_{i} \,.
\end{equation}

There are many interesting details of how the magnitude of the baryon
asymmetry may be correlated with some other physical quantity in
a particular model.
For instance, the $SU(5)$ model with the minimal Higgs structure
yields too small a value of $n_{B}/n_{\gamma }$. On the other hand,
the baryon asymmetry is correlated with the neutrino
mass in $SO(10)$ models \cite{fyy81}:
the smaller the neutrino mass is, the smaller the baryon asymmetry is.
This can be utilized to constrain a finite, but a small neutrino mass.
I shall however omit discussion of these subjects in this introductory
lecture.

It is now appropriate to discuss difficulty associated with the
GUT baryogenesis and how to overcome it.
First, as I already mentioned above, there may be baryon number
annihilation via electroweak processes at finite temperatures.
One can always excuse for the GUT scenario
that the electroweak baryon nonconservation
keeps intact $B - L$, so if $B - L$ is created at the GUT epoch, later
evolution only redistributes $B$ and $L$, but never annihilates 
the baryon number;
\begin{equation}
B_{{\rm final}} = c\cdot (B - L)_{{\rm initial}} \,, 
\end{equation}
with 
\( \:
L_{{\rm final}} = (1-c)\cdot (B - L)_{{\rm initial}} \,.
\: \)
Here c is a calculable number of order unity.

One has to be careful however to avoid a large lepton violation at
intermediate temperature scales that may potentially dissipate
away the baryon number with 
\( \:
\Delta B = \Delta L = 0 \,, 
\: \)
when combined with electroweak $B$ nonconservation.
Allowed range of this interaction, accordingly constraint on 
the neutrino mass via the seesaw type of neutrino mass generation may
be estimated by introducing a generic type of $\Delta L = 2$ interaction;
\begin{equation}
{\cal L}_{\Delta L \neq 0} = \frac{m_{\nu }}{v^{2}}\,l_{L}l_{L}\varphi 
\varphi + ({\rm h.c.}) \,,
\end{equation}
with $l_{L}$ and $\varphi $ the lepton and the Higgs doublet, respectively
and $v \simeq 250\,$GeV.
Requirement of harmless lepton violation,
\( \:
\Gamma _{\Delta L \neq 0} < H \,, 
\: \)
leads to the upper bound on the neutrino mass \cite{harvey-turner};
\begin{equation}
m_{\nu } < 4\times 10^{-3}\,{\rm eV}\,(T_{B-L}/10^{16}\,{\rm GeV})^{-1/2}
\,, 
\end{equation}
with $T_{B-L}$ the scale of $B-L$ generation at higher temperatures.
This constraint is always applied to the lightest neutrino species, but
when the neutrino mixing is large, it is also applied to the heaviest
neutrino species.

The second problem with the GUT baryogenesis is more serious;
a possible overproduction
of gravitino and associated low reheating temperature after inflation.
The gravitino is the superpartner of graviton in supergravity theories.
It is a spin $3/2$ particle and couples with
ordinary matter field with the gravitational strength.
Moreover, one usually associates supersymmetry breaking scale with 
the electroweak scale in order to ease the hierarchy problem, thus
the mass of the gravitino $m_{3/2} = O[\,{\rm TeV}\,]$.
This has a consequence potentially very serious, because the decay rate
of the gravitino is given by
\begin{equation}
\Gamma _{3/2} \approx \frac{m_{3/2}^{3}}{m_{{\rm pl}}^{2}}
\sim (10^{5}\,{\rm sec})^{-1}\cdot (\frac{m_{3/2}}{{\rm TeV}})^{3}\,,
\end{equation}
the lifetime being close to the epoch of nucleosynthesis.
If the gravitino abundance is larger than of order $10^{-10}$ relative
to the entropy density, the successful nucleosynthesis would be destroyed.

For a short while inflation was considered to save this potential disaster
by diluting away the gravitino abundance.
But subsequently it has been recognized \cite{gravitino problem}
that regeneration of gravitinos
after inflation severely constrains the maximally allowed reheating
temperature. 
The point of this argument is that  gravitino pairs may be produced
from ordinary particles, whose cross section
is calculable and of order $1/m_{{\rm pl}}^{2}$, giving the abundance
of order 
\begin{equation}
\frac{n_{3/2}}{s} = O[\,10^{-2}\,]\,\frac{T}{m_{{\rm pl}}}\,,
\end{equation}
in terms of the reheat temperature $T$ after inflation.
For successful nucleosynthesis one gets the constraint on 
the reheat temperature
\begin{equation}
T < O[\,10^{10} - 10^{11}\,]\,{\rm GeV}
\end{equation}
from 
\( \:
\frac{n_{3/2}}{s} < O[\,10^{-10} - 10^{-11}\,] \,.
\: \)
This temperature is too low to create the baryon asymmetry 
by the $X$ boson decay.
It thus appeared that supergravity models arising as the field theory limit of
presumably the ultimate theory of superstring give too low a reheat
temperature incompatible with the GUT baryogenesis.

I shall discuss in the following sections how a correct theory of reheating
after inflation may provide a high temperature phase suitable to
the baryon generation without overproduction of gravitinos.

\section{Thermal history after inflation}
How the hot big bang is started after inflation is a fascinating subject 
that can be discussed independently of baryogenesis.
But since the subject directly addresses the origin of entropy in
our universe at earliest times of evolution, 
one may very naturally entertain the possibility
that the two basic quantities in cosmology, 
the baryon number and the entropy are both created roughly at the same time.
Moreover, association of the origin of the cosmic entropy with inflation
sets an ideal theoretical framework, because inflation dilutes away
everything in our observable part of the universe: one must explain
the origin of entropy starting from empty space, except the coherent
inflaton oscillation around the minimum of inflaton potential.

For quite some time the theory of particle production due to coherent
and homogeneous field oscillation was based on a naive picture of
lowest order perturbation \cite{reheat-original}.
The oscillating inflaton field, denoted here by $\xi (t)$, is regarded
in this naive picture as an aggregate of condensed bosons at rest,
and these bosons are assumed to decay stochastically according to
the rate given by the Born approximation.
Thus equating the decay rate with the Hubble rate 
\( \:
\gamma _{\xi } = H \,, 
\: \)
together with assumption of the instantaneous reheating, leads to
\begin{equation}
T_{B} \sim 0.1\,\sqrt{\,\gamma _{\xi }m_{{\rm pl}}\,}
\sim 10^{14}\,{\rm GeV}\cdot g\cdot \frac{m_{\xi }}{10^{13}\,{\rm GeV}} \,.
\end{equation}
The Born rate for massless particle decay,
\( \:
\xi \rightarrow \varphi \varphi \,, 
\: \)
\begin{equation}
\gamma _{\xi } = \frac{g^{2}}{32\pi }\,m_{\xi }
\end{equation}
was used in this estimate.

This picture is valid when the amplitude of oscillation and the
coupling $g$ to matter field is small enough, but is grossly wrong
if this condition is violated, for instance when the amplitude of
oscillation is large.
As will be explained shortly, 
there exist an infinitely many bands of instability that
may contribute to the inflaton decay, only one of which,
the lowest band, when restricted to the small amplitude limit, 
is identified with the Born decay rate.
This phenomenon of instability is known as the parametric resonance
under periodic potential \cite{landau-lifschitz m},\cite{coddington}.

A systematic method to understand particle production and associated inflaton
decay can be formulated \cite{mine95-1},\cite{reheating-rev-mine} 
in the Schr$\stackrel{..}{{\rm o}}$dinger picture of quantum field theory
. (For other approaches and other aspects of this problem, see ref
\cite{reheating parametric},\cite{linde et al 94},\cite{holman 95},
\cite{brandenberger et al}.)
Quantum bose fields that couple to the inflaton field are treated
as a quantum operator (in the Heisenberg picture), 
but the inflaton field is regarded as classical in this approach,
although back reaction against the inflaton oscillation is also considered.
In the Schr$\stackrel{..}{{\rm o}}$dinger picture the state vector 
describing behavior of quantum field coupled to the inflaton 
is given by a direct product of
state vectors $|\psi (t) \rangle _{\vec{k}}$ 
of independent spatial Fourier modes.
During the time interval $\Delta t$ that obeys
\begin{equation}
1/m_{\xi } \ll \Delta t \ll 1/H \,, 
\end{equation}
one may assume exactly periodic inflaton oscillation 
with a periodically varying frequency,
\( \:
\omega _{\vec{k}}^{2}(t) \,,
\: \)
containing the oscillating function $\xi (t)$.
For longer time scales,
\( \:
\Delta t \gg 1/H \,, 
\: \)
the amplitude of oscillation is taken to adiabatically change.

A salient feature of this approach is that one may solve the quantum
state with a Gaussian ansatz;
\begin{equation}
\langle q_{k} |\psi (t) \rangle_{\vec{k}} = 
\frac{1}{\sqrt{u_{k}}}\,\exp [\,\frac{i}{2}\frac{\dot{u}_{k}}{u_{k}}\,
q_{k}^{2}\,] \,,
\end{equation}
where $u_{k}(t)$ is shown to obey the classical oscillator equation;
\begin{eqnarray}
[\,\frac{d^{2}}{dt^{2}} + \omega _{\vec{k}}^{2}(t)\,]\,u_{k} = 0
\,.
\end{eqnarray}
The initial condition for this classical equation is determined by
the choice of an initial quantum state.
The simplest, and a reasonable choice is the ground state with respect
to some reference frequency, most naturally the frequency at the onset of
inflaton oscillation $\omega _{k}(0)$. 
This is because after inflation the state is
essentially devoid of matter.
Under this circumstance the initial condition is
\begin{equation}
u_{k}(0) = (\frac{\omega _{0}}{\pi })^{-1/2} \,, \hspace{0.5cm} 
\frac{\dot{u}_{k}}{u_{k}}(0) = i\omega _{0} \,, 
\end{equation}
with $\omega _{0}$ the reference frequency, and one may further simplify
the wave function in terms of $|u_{k}(t)|$ alone \cite{mine95-1}.

The most important consequence \cite{mine95-1},\cite{reheating-rev-mine} 
of this formalism is that it gives a rationale
to introduce a coarse grained density matrix $\rho ^{(D)}$
which has a classical probability distribution.
The coarse graining here is defined by a short time average over the time
scale of order a few oscillation periods.
This seems a reasonable way to extract global behavior of the quantum
system ignoring fine details of the quantum state.
This makes it possible to replace the quantum density matrix by
the time-averaged diagonal part 
\( \:
\langle n|\rho ^{(D)}|n \rangle
\: \)
of the density matrix in a convenient 
base such as the Fock base of frequency $\omega _{0}$:
\begin{equation}
\langle n|\rho ^{(D)}|n \rangle = \overline{\langle n|\psi (t) \rangle
\langle \psi (t)|n \rangle} \,, 
\end{equation}
where $|n\rangle $ is the $n-$th level of field oscillators and
the overline represents the short time average.
After the coarse graining a finite entropy may be assigned;
\begin{equation}
-\,{\rm tr}\,\rho ^{(D)}\ln \rho ^{(D)} > 0 \,.
\end{equation}

From this reduced density matrix one computes various physical quantities.
For instance, the produced particle number in each mode is given by
\begin{eqnarray}
\langle N_{k} \rangle &=& \langle \,\frac{1}{\omega _{0}}
(\frac{1}{2}\, p_{k}^{2} + \frac{1}{2}\, \omega _{0}^{2}q_{k}^{2})
- \frac{1}{2}\, \rangle = {\rm tr}\,(\rho_{k} ^{(D)}\,a_{k}^{\dag }a_{k})
 \nonumber \\
&=&
\frac{4(\,\omega |u_{k}|^{2} - \pi \,)^{2} + 
(\,\frac{d|u_{k}|^{2}}{dt}\,)^{2}}{16\pi\, \omega\, |u_{k}|^{2}} \,, 
\end{eqnarray}
with $q_{k} \,, p_{k}$ oscillator coordinates.
Suitable time average is understood here.
As $t\rightarrow \infty $, 
\begin{equation}
\langle N_{k} \rangle \rightarrow e^{\lambda _{k}m_{\xi }t} \times 
{\rm (polynomial\; in}\; t)
\end{equation}
for the momentum $k$ in the instability band.

Two types of matter coupling have been considered;
\begin{equation}
V_{\xi } = \frac{1}{2}\, g^{2}\,\xi ^{2}\varphi ^{2} + 
\frac{1}{2}\, cgm_{\xi }\,\xi \varphi ^{2} \,, 
\end{equation}
with $c$ a constant of order unity and $m_{\xi }$ the mass of the inflaton.
As a model of inflation we consider the simplest chaotic type inflation
\cite{linde83} with parameters;
\( \:
m_{\xi } \sim 10^{13}\,
\: \)
GeV consistent with COBE anisotropy, 
and the initial amplitude of order the Planck scale,
\begin{equation}
\xi _{0} \sim \sqrt{\frac{3}{4\pi }}\,m_{{\rm pl}} \,.
\end{equation}
The dimensionless coupling $g$ is taken arbitrary at the moment.

The classical mode equation is then 
\begin{eqnarray}
&&
\frac{d^{2}u}{dz^{2}} + [\,h - 2\theta \cos (4z) - 4c\sqrt{\theta }
\sin (2z)\,]u = 0 \,, 
\label{classical mode eq} \\
&&
h = 4\frac{\vec{k}^{2} + m^{2}}{m_{\xi }^{2}} + 2\theta \,, \hspace{0.5cm} 
\theta = \frac{g^{2}\xi_{0} ^{2}}{m_{\xi }^{2}} \,, \hspace{0.5cm} 
z = \frac{m_{\xi }t}{2} \,,
\end{eqnarray}
where $\xi _{0}$ is the amplitude of inflaton oscillation;
\begin{equation}
\xi (t) = -\,\xi _{0}\,\sin (m_{\xi }t)  \,.
\end{equation}
The criterion of large or small amplitude is thus given by the magnitude
of $\theta $. In the very small $(\theta \ll 1)$ or in the very large
$(\theta \gg 1)$ amplitude limit the classical mode equation 
(\ref{classical mode eq}) reduces
to the Mathieu equation \cite{mathieu eq} with a single oscillating term.
In both these limits we developed analytic formulas suitable for
detailed analysis of the reheating problem and related problems, too
\cite{mine95-1},\cite{fkyy95-1}.

The structure of instability bands is as follows.
Each band is labeled by an integer $n = 1 \,, 2\,, 3 \cdots $, and
goes to $h \rightarrow n^{2}$ in the small amplitude limit, $\xi _{0}
\rightarrow 0$. The band width in the small amplitude region is
\begin{equation}
\Delta h_{n} = \frac{\theta ^{n/2}c_{n}}{2^{2n - 1}[(n-1)!]^{2}} \,, 
\end{equation}
at a fixed $\theta $, with $c_{n}$ some function of $c$ 
\cite{mine95-1},\cite{mine96-1}.
For instance,
\( \:
c_{n} = 2^{n}
\: \)
with the Yukawa coupling
\( \:
\frac{1}{2}\, gm_{\xi }\,\xi \varphi ^{2}
\: \)
alone.
The instability bands are thus indeed very narrow in the small amplitude
region.
But as $\theta $ increases, the bands become broad, and 
for $h < 2\theta $ most of the $(\theta \,, h)$
parameter space is covered by instability
bands except bounding narrow stability bands.
For parameters $(h \,, \theta) $ or corresponding $(k \,, \xi _{0})$ within
an instability band the classical solution exhibits exponential growth;
\begin{equation}
u \rightarrow e^{\lambda m_{\xi }t} \times P(t)\,,
\end{equation}
with $P(t)$ a periodic function.
This has the important consequence of exponential particle production rate
unless the back reaction stops it.

In the small amplitude limit exact results for the growth rate
$\lambda $ and mode sum are available.
After the coarse graining the initial state decays according to
\( \:
e^{-\,\Gamma Vt}
\: \)
where
\begin{equation}
\Gamma = \sum_{n}\,\Gamma _{n} \,, \hspace{0.3cm}
\Gamma _{n} = \frac{m_{\xi }}{2V}\sum_{\vec{k}\, \in \,n-{\rm th \; band}}
\! \lambda _{\vec{k}}\,.
\end{equation}
$\Gamma _{n}$ is the decay rate per unit volume of the $n-$th band,
which is computed  \cite{mine95-1},\cite{mine96-1} as
\begin{equation}
\Gamma _{n} = \frac{m_{\xi }^{4}}{64\pi }\,\sqrt{\,1 - \frac{4m^{2}}
{n^{2}m_{\xi }^{2}}\,}\,(\Delta h_{n})^{2} \,.
\end{equation}
It was also shown recently  \cite{mine96-1}
that the small amplitude result can be understood 
by familiar perturbation theory.
The point is that the mode-summed rate grows with the amplitude as
\( \:
\propto \xi _{0}^{2n} \propto (n_{\xi })^{n}
\: \)
where $n_{\xi } = \frac{1}{2}\, m_{\xi }\,\xi _{0}^{2}$ is the number 
density of condensed inflatons. The decay rate precisely
coincides with the zero-momentum limit of $n$ to 2 body process
\( \:
n\,\xi \rightarrow \varphi\, \varphi \,,
\: \)
worked out using the ordinary Feynman rule.
In particular, the decay rate of the first band with $c = 1$ is given by
\begin{equation}
\Gamma _{1} = \frac{g^{2}m_{\xi }^{2}\,\xi _{0}^{2}}{64\pi }\,
\sqrt{\,1 - \frac{4m^{2}}{m_{\xi }^{2}}\,} \,.
\end{equation}
This exactly coincides with the one-particle decay rate $\gamma _{\xi }$
for $\xi \rightarrow \varphi \varphi $, when divided by the inflaton
number density $\frac{1}{2}\, m_{\xi }\,\xi _{0}^{2}$.

It is important for many interesting applications to work out analytic
formulas in the large $\theta $ region. This seems a fomidable task in view of
that non-perturbative analysis is involved.
A remarkable result \cite{fkyy95-1} is that in the functional 
Schr$\stackrel{..}{{\rm o}}$dinger picture one can solve the fundamental
quantity $u_{k}(t)$, which becomes rigorous in the region,
\( \:
\theta \gg 1
\: \) with
\( \:
|h - 2\theta | \ll \theta \,.
\: \)
Furthermore this is precisely the parameter region of prime importance
when one considers cosmological evolution, as will be discussed
shortly.
In the rest of this lecture 
we shall neglect the mass $m$ of quantum bose field considering
only those of $m \ll m_{\xi }$.

Cosmological evolution changes both the momentum and the inflaton
amplitude according to
\begin{equation}
k \propto 1/a \,, \hspace{0.3cm} \xi \propto 1/a^{3/2} \,.
\end{equation}
Taking in the large $\theta $ region the leading variation alone
gives the approximate rule;
\( \:
\Delta h \sim 2\,\Delta \theta \,.
\: \)
Thus parameters of the dominant contribution move parallel to the
$h = 2\theta $ line with cosmological evolution.
Moreover, the largest particle production with the largest rate $\lambda $ 
occurs in the deepest region within instability bands of
the Mathieu equation (the original
two terms of oscillations reduce to one term for $\theta \gg 1$).
This implies that the most dominant region is along $ h = 2\theta $ 
within some width $\delta  = \delta (h - 2\theta )$.

The analytic result derived in ref \cite{fkyy95-1} is summarized as
\begin{eqnarray}
&&
\lambda = \frac{1}{\pi }\,\ln (\sqrt{x} + \sqrt{x-1}) \,, \\
&&
x = (1 + e^{-\pi \delta /(4\sqrt{\theta })})\,\cos^{2}\psi \,, \\
&& \hspace*{-0.5cm}
\psi = \frac{\pi ^{2}}{4}\sqrt{\theta } + \frac{\delta }{4\sqrt{\theta }}
\ln (\frac{\pi \theta ^{1/4}}{\sqrt{2}}) +
\frac{1}{2} \Im \,\ln [\,
\frac{\Gamma \left(\frac{1}{2} - i\delta  /(8\sqrt{\theta })\right)}
{\Gamma \left(\frac{1}{2} + i\delta  /(8\sqrt{\theta })\right)}\,]
\,. 
\end{eqnarray}
The instability region is characterized by $x > 1$, while the stability
region by $x < 1$.
These two regions alternate roughly with equal band width of
\( \:
\Delta \theta \approx \frac{4}{\pi }\sqrt{\theta }
\: \)
around along $h = 2\theta $.
For $\delta  \ll \sqrt{\theta }$, or $k \ll \sqrt{gm_{\xi }\xi _{0}}$,
it approximately follows that
\begin{equation}
\lambda = \frac{1}{\pi }\,\ln (\,\sqrt{2}\,|\cos \psi | + 
\sqrt{\,|\cos (2\psi )|\,}\,) \,, \hspace{0.5cm} 
\psi = \frac{\pi ^{2}}{4}\sqrt{\theta } \,.
\end{equation}
It can be readily shown that the maximal $\lambda $ along
$h = 2\theta $ is
\begin{equation}
\lambda _{{\rm max}} = \frac{1}{\pi }\ln (\sqrt{2} + 1) \simeq 0.28
\end{equation}
and the average in the instability band is
\( \:
\bar{\lambda } \simeq 0.22 \,.
\: \)
It is evident that in the large $\theta $ region the growth rate
never diminishes, always with a sizable constant $\lambda $.
Thus the exponent of particle production rate grows roughly in the time 
interval of
$1/\bar{\lambda }\,(\approx 5)\,\times $ oscillation period $1/m_{\xi }$.

Particle production is halted by the back reaction.
This problem is studied by solving a coupled system of the inflaton
and the radiation energy densities in the expanding universe
\cite{fkyy95-2},
\begin{eqnarray}
&&
\frac{d\rho _{\xi }}{dt} + 3H\rho _{\xi } = -\,N\frac{d}{dt}\,
\langle \rho _{\varphi } \rangle\,, \\
&&
\frac{d\rho _{r }}{dt} + 4H\rho _{r} = N\frac{d}{dt}\,
\langle \rho _{\varphi } \rangle\,, \\
&&
H^{2} = \frac{8\pi }{3}G\,(\rho _{\xi } + \rho _{r}) \,,
\end{eqnarray}
where $N$ is the number of boson species contributing to parametric
resonance effects, roughly with the coupling $g$.

The idea behind this evolution equation is that irrespective of
interaction among created particles towards thermalization,
the energy balance between the inflaton and radiation should hold
at any instant of time because of energy conservation.
If one separately checks that thermalization is realized, the radiation
energy density $\rho _{r}$ can be used to estimate the temperature
in equilibrium;
\begin{equation}
T = (\frac{30}{\pi ^{2}N'}\,\rho _{r})^{1/4}\,.
\end{equation}
We allow the possibility that there might be more particle species
participating in equilibration than the number of created boson
species $N$: for instance, in supersymmetric theories of $N$ boson species
there are roughly equal number of fermion species that may be produced
from the bosons by secondary processes, giving $N' \approx 2N$.
The system of evolution equation can be extended to include the mass
variation by
\begin{equation}
\frac{dm_{\xi }^{2}}{dt} = N\cdot g^{2}\,\frac{d}{dt}
\langle \varphi ^{2} \rangle \,.
\end{equation}

It might be instructive to give derivation of this evolution equation
in view of that the meaning of energy densities in the presence of
interaction terms is not clear.
We define the inflaton energy density including the mass variation;
\begin{equation}
\rho _{\xi } = \frac{1}{2}\, \dot{\xi }^{2} + \frac{1}{2}\, 
m_{\xi }^{2}(t)\,\xi ^{2} \,, 
\end{equation}
with the time dependent part of the mass given by
\( \:
g^{2}\sum_{\varphi }\,\langle \varphi ^{2} \rangle
\,.
\: \)
From the $\xi $ equation of motion,
\begin{eqnarray}
\stackrel{..}{\xi } + 3H\dot{\xi } + m_{\xi }^{2}(t)\,\xi = 
-\,\langle\, \frac{\partial V_{{\rm Y}}}{\partial \xi }\, \rangle \,, 
\end{eqnarray}
where $\langle A \rangle \equiv {\rm tr}\,(A\rho )$ 
($\rho =$ density matrix) and 
\( \:
V_{{\rm Y}} = \frac{1}{2}\,gm_{\xi }\xi\, \sum_{\varphi } \varphi ^{2}
\: \), 
one derives for evolution of the $\xi $ energy density;
\begin{eqnarray}
\dot{\rho }_{\xi } + 3H\dot{\xi }^{2} = 
-\,\langle\, \dot{\xi }\, \frac{\partial V_{{\rm Y}}}{\partial \xi } \,\rangle 
+ \frac{g^{2}}{2}\,\xi ^{2}\,\frac{d}{dt}\sum_{\varphi }
\,\langle \varphi ^{2} \rangle
\,.
\end{eqnarray}
Note that we included the quartic interaction term 
\( \:
\frac{1}{2}\, g^{2}\xi ^{2}\,\sum_{\varphi }\,\langle \varphi ^{2} \rangle
\: \)
in the time variant mass term $m_{\xi }^{2}(t)$.
The left hand side becomes $\dot{\rho }_{\xi } + 3H\rho _{\xi }$
after time average over the oscillation period, since
\begin{equation}
\overline{\dot{\xi }^{2}} = \overline{m_{\xi }^{2}(t)\,\xi ^{2}} =
\overline{\rho _{\xi }} \,,
\end{equation}
under the assumption of slow $\xi $ variation.

On the other hand, it can be shown 
using the time evolution equation for the density matrix,
\begin{equation}
\dot{\rho } = -\,i[\,H \,, \rho \,] = -i\,[\,\sum_{\varphi }
H_{\varphi  } + V_{{\rm Y} } + V_{4} \,, \rho \,] \,, 
\end{equation}
that
\begin{eqnarray}
\langle \,\dot{\xi }\,\frac{\partial V_{{\rm Y}}}{\partial \xi } \, \rangle
 &=& {\rm tr}\,(\,\frac{\partial V_{{\rm Y}}}{\partial t}\,\rho \,) =
 \frac{d}{dt}\,\langle V_{{\rm Y}} \rangle - {\rm tr}\,(V_{{\rm Y}}\,
 \dot{\rho }) \nonumber \\
&=&
 \frac{d}{dt}\,\langle V_{{\rm Y}} \rangle + \frac{d}{dt}\,
 \sum_{\varphi }\, \langle H_{\varphi } \rangle +
\frac{d}{dt}\,\langle V_{4} \rangle - 
\langle \frac{\;\partial V_{4}}{\partial t} \, \rangle
\,, 
\end{eqnarray}
where 
\begin{equation}
H_{\varphi } = \frac{1}{2}\, \dot{\varphi }^{2} + \frac{1}{2}\, 
(\nabla \varphi)^{2}
\end{equation}
is the free Hamiltonian density of created boson
$\varphi $ and 
\( \:
V_{4} = \frac{1}{2}\, g^{2}\xi ^{2}\,\sum_{\varphi }\varphi ^{2}
\: \).
The short time average over one oscillation period of 
$\overline{\langle V_{{\rm Y}} \rangle}$ vanishes, since this quantity
has a linear dependence on $\xi $.
Thus one has
\begin{eqnarray}
&&
-\,\langle\, \dot{\xi }\, \frac{\partial V_{{\rm Y}}}{\partial \xi } \,\rangle 
+ \frac{g^{2}}{2}\,\xi ^{2}\,\frac{d}{dt}\sum_{\varphi }
\,\langle \varphi ^{2} \rangle \nonumber \\
&& \hspace*{0.5cm} =
-\,\frac{d}{dt}\,\sum_{\varphi }\,\langle H_{\varphi } \rangle 
- \frac{g^{2}}{2}\,\sum_{\varphi }\,\langle \varphi ^{2} \rangle\,
\frac{d\xi ^{2}}{dt} +
\langle \frac{\;\partial V_{4}}{\partial t} \, \rangle \,,
\end{eqnarray}
leading to
\begin{eqnarray}
&&
\dot{\rho }_{\xi } + 3H\rho _{\xi } = -\,
\frac{d}{dt}\,\sum_{\varphi }\,\langle H_{\varphi } \rangle 
\,, \\
&&
\dot{\rho }_{r} + 4H\rho _{r} = 
\frac{d}{dt}\,\sum_{\varphi }\,\langle H_{\varphi } \rangle \,.
\end{eqnarray}
The second equation simply follows from energy conservation
of the combined system of $\xi $ and radiation.

It is now appropriate to discuss the thermal history after inflation,
namely after the inflaton commences oscillation at $t \approx  \frac{2}{3}
\frac{1}{m_{\xi }}$. The result should be compared with the naive
estimate of the reheat temperature, 
\( \:
T_{B} \sim 0.1\sqrt{\,\gamma _{\xi }\,m_{{\rm pl}}\,}
\sim 10^{-2}\,g\sqrt{\,m_{\xi }\,m_{{\rm pl}}\,} \,.
\: \)

Particle production is significant only when the exponent in the rate
is appreciable, but once it becomes of order unity, the production rate
becomes accelerated. Indeed, when the exponent becomes of order 100,
a catastrophic particle production occurs and back reaction immediately
stops particle production.
The time when this happens can be estimated by
\begin{equation}
t = t_{d} \approx O[100]\,/(\bar{\lambda }m_{\xi }) \,.
\end{equation}

This is an abrupt change, as numerically checked by solving the time
evolution equation \cite{fkyy95-2}. 
Prior to this time the inflaton density varies
according to the $\xi-$matter dominance,
\( \:
\rho _{\xi } = \frac{1}{2}\, m_{\xi }^{2}\,\xi ^{2} \propto 1/t^{2} \,.
\: \)
Combined with the time of catastrophic decay $t_{d}$ above, the abrupt change
occurs at
\begin{equation}
\frac{\xi}{\xi _{0}} \approx 
O[\,10^{-2}\,]\,\frac{2}{3}\,\bar{\lambda } \,.
\end{equation}

A typical energy of produced particles at this time, prior to
thermalization, is of order
\begin{equation}
\bar{E} = O[\sqrt{gm_{\xi }\,\xi }] = O[\,0.3\,]\,
\sqrt{g\bar{\lambda }m_{\xi }\,m_{{\rm pl}}}\,, 
\end{equation}
with $\xi $ given as above.
Using this energy, one estimates the rate of two body reactions
among created particles, and
the corresponding ratio of this to the Hubble rate,
\begin{eqnarray}
\frac{\Gamma }{H} &\approx& \frac{\alpha _{s}/\bar{E}^{2}\cdot Nn_{\varphi }}
{\sqrt{\rho _{\xi } + \rho _{r}}/m_{{\rm pl}}}
 \,, \nonumber \\
 &\approx &
\alpha _{s}\frac{m_{{\rm pl}}}{\bar{E}}(\frac{\rho _{r}}{\bar{E}^{4}})^{1/2}\,
\sqrt{\frac{\rho _{r}/\rho _{\xi }}{1 + \rho _{r}/\rho _{\xi }}} \,,
\end{eqnarray}
with $\alpha _{s}$ a typical coupling involving ordinary particles.
What usually happens \cite{fkyy95-2} is that 
\begin{equation}
\rho _{r} \approx \rho _{\xi } \,, \hspace{0.5cm} 
\frac{\rho _{r}}{\bar{E}^{4}}  \gg 1 \,, 
\end{equation}
hence 
\( \:
\Gamma \gg H \,.
\: \)
Thus two-body reactions frequently take place.
Even multi-particle reactions occur and one may conclude that
thermalization takes place immediately after the catastrophic particle
production.
Under this circumstance the reheat temperature right after the
explosive particle production is
\begin{eqnarray}
T_{R} \approx O[10^{-2}]\,\sqrt{\bar{\lambda }m_{\xi }m_{{\rm pl}}} \,.
\end{eqnarray}
With 
\( \:
m_{\xi } \sim 10^{13}\,{\rm GeV} \,, \;
\bar{\lambda } \approx 0.2 \,, 
\: \)
this reheat temperature is close to the GUT scale.
It is possible that at least the Higgs $X-$boson of mass $\sim 10^{14}\,$
GeV can copiously be produced.
In any case $T_{R} \gg  T_{B}$ for a small coupling $g$.
More precise computation has been done by numerically integrating
time evolution equation \cite{fkyy95-2}.

The entire thermal history may in general be fairly complicated.
Even if the radiation dominance is realized at the early epoch simultaneous
to the catastrophic production,
there exists a residual inflaton field after the catastrophic stage:
\begin{equation}
\theta \approx O[10^{-4}]\,\frac{1}{3\pi }(\frac{g\bar{\lambda }m_{{\rm pl}}}
{m_{\xi }})^{2} \,.
\end{equation}
If this value is still large, $\gg 1$, there may be a second explosive
decay. On the other hand, if this value is moderately small, 
but not small enough such that the naive perturbative analysis is 
no longer valid,
then there may be gradual particle production continuously down to
$\theta \ll 1$. This region is difficult to analyze, but under study
currently.

The ultimate end point of the inflaton decay is the decay in the first
band, namely the Born decay 
when the inflaton amplitude becomes very small.
This is because only the first band appreciably contributes
with very small $\theta $,
and the Born decay rate always satisfys $\gamma _{\xi } > H$
at very late times. 
Under the assumption that intermediate amplitude decay is not
significant, one can estimate entropy creation at the Born decay.
According to our numerical integration \cite{fkyy95-2} 
the entropy production by the Born decay
is significant for the coupling range $g < O[10^{-3}]$, but negligible
for $g > O[10^{-2}]$.
In all cases numerically checked \cite{fkyy95-2}, 
the final temperature at the end of
the complete inflaton decay is given approximately by the estimate
due to the Born formula $T_{B}$.
This does not mean that effects of parametric resonance can be forgotten.
High energy processes that may occur in the temperature range,
$T_{R} > T > T_{B}$, have to be reconsidered.
As such, the baryogenesis and the gravitino production are of prime
importance.

\section{Gravitino abundance}
The old formula of the gravitino abundance
\( \:
n_{3/2}/s \sim 10^{-2}\cdot T_{B}/m_{{\rm pl}}
\: \)
with $s$ the entropy density is not valid since the thermal history
right after the catastrophic particle production is complicated in
some region of the coupling $g$, and $T_{B}$ is not a good measure to
characterize the thermal history.
The truth is that no single temperature represents the state after
inflation.
In order to estimate the gravitino abundance with non-trivial thermal
history after inflation, it is necessary to follow time evolution of
the gravitino number density $n_{3/2}$ \cite{fkyy95-2},
\begin{equation}
\frac{dn_{3/2}}{dt} + 3H\,n_{3/2} = \langle \Sigma v \rangle\,
n_{\varphi }^{2} \,,
\end{equation}
where $n_{\varphi }$ is the thermal number density of one species of
created bosons. The cross section $\langle \Sigma v \rangle$ of
gravitino production $\varphi \,\varphi \rightarrow g_{3/2}\:g_{3/2}$
has been computed, and roughly 
\( \:
\langle \Sigma v \rangle \sim 250/m_{{\rm pl}}^{2}
\: \) \cite{gravitino problem} 
unless the gravitino is lightest supersymmetric particle (LSP).
Destructive term has been neglected in the evolution equation above, 
which is justified for the gravitino mass larger than $O[1]\,$keV
\cite{stable gravitino}.

Both possibilities of stable and unstable gravitino remain viable
\cite{fkyy95-2}. 
Let us only mention a possibility
of the gravitino dominated universe at the present epoch 
\cite{stable gravitino}.
With the initial temperature $T_{R} > 2\times 10^{14}\,{\rm GeV}$ imposed
to give a favorable situation for GUT baryogenesis,
there is a region of parameters for the closure density of gravitino
dominated universe if $m_{3/2} = 0.1 - 10\, {\rm GeV}$.
The basic reason this becomes possible is that the initial large gravitino
yield created right after the catastrophic particle production
is much diluted via the late phase of Born decay.
This is again a reflection of the large disparity of the two temperatures;
\( \:
T_{R} \gg T_{B} \,.
\: \)
Of course, it remains to demonstrate a sizable baryon to photon ratio.
But things are not bad: there is an epoch immediately before the catastrophic
particle production in which non-equilibrium environment necessary for
baryon generation exists, and moreover the observed baryon to photon ratio
is of order $10^{-10}$, allowing some amount of dilution in later epochs.

\def\lromn#1{\uppercase\expandafter{\romannumeral#1}}
\def\romn#1{\romannumeral#1}


\begin{thebibliography}{99}

\bibitem{inflation-rev}
For a review,
A. Linde, {\em Particle Physics and Inflationary Cosmology},
(Harwood, Chur, Switzerland, 1990).

\bibitem{3kfluc-obs}
G. F. Smoot, 
in {\em Cosmological Constant and the Evolution
of the Universe}, ed. by K. Sato, T. Suginohara, and
N. Sugiyama (Universal Academy Press, Tokyo, Japan, 1996).

\bibitem{nucleosyn-rev}
A.M. Boesgaard, and G. Steigman, {\sl Ann.\ Rev.\ Astron.\ Astro.\ },
{\bf 23}, 319(1985).

\bibitem{sakharov}
A.D. Sakharov, {\sl JETP Lett.\ }{\bf 5}, 24(1967).

\bibitem{my78} 
M. Yoshimura, {\sl Phys.\ Rev.\ Lett.\ }{\bf 41}, 281(1978);{\bf 42},746(E)
(1979);
M. Yoshimura, {\sl Phys.\ Lett.\ }{\bf B88}, 294(1979).

\bibitem{ttez} 
D. Toussaint, S.B. Treiman, F. Wilczek and A. Zee, 
{\sl Phys.\ Rev. }{\bf D19}, 1036(1979).

\bibitem{wein79} 
S. Weinberg, {\sl Phys.\ Rev.\ Lett.\ }{\bf 42},850(1979).

\bibitem{dimosuss} 
S. Dimopoulos and L. Susskind, {\sl Phys.\ Rev. }{\bf D18}, 4500(1978).

\bibitem{basym-review-mine}
M. Yoshimura, 
{\em Cosmological Baryon Production and Related Topics},
in Proceedings of the Fourth Kyoto Summer Institute on Grand Unified Theories 
and Related Topics, ed. by M. Konuma and T. Maskawa, 
(World Scientific Pub., Singapore. 1981).

\bibitem{basym-review-kolbt}
E.W. Kolb and M.S. Turner,{\sl Annu.\ Rev.\ Nucl.\ Part.\ Sci.\ }
{\bf 23}, 645(1983).

\bibitem{gut-rev-langacker}
For a review, P. Langacker, {\sl Phys.\ Rep.\ }{\bf 72}, 185(1981).

\bibitem{thooft-76}
t' Hooft, {\sl Phys.\ Rev.\ Lett.\ }{\bf 37}, 8(1976).

\bibitem{krs-ew-bgen}
V.A. Kuzmin, V.A. Rubakov and M.E. Shaposhnikov, {\sl Phys.\ Lett.\ }
{\bf B155}, 36(1985).

\bibitem{sphaleron}
F.R. Klinkhamer and N.S. Manton, {\sl Phys.\ Rev. }{\bf D30}, 2212(1984).

\bibitem{ew-bgeneration-review}
For a review, A.G. Cohen, D.B. Kaplan, and A.E. Nelson, \\
{\em Progress in Electroweak Baryogenesis},
{\sl Annu.\ Rev.\ Nucl.\ Part.\ Sci.\ }{\bf 43}, 27(1993).

\bibitem{farrar-shapo}
G.R. Farrar and M.E. Shaposhnikov, {\sl Phys.\ Rev. }{\bf D50}, 774(1994).

\bibitem{basym-standard-th}
M.B. Gavela, P. Hernandez, J. Orloff, O. P\`ene, and C. Quimbay,
{\sl Nucl.\ Phys.\ }{\bf B430}, 382(1994);

P. Huet and E. Sather, {\sl Phys.\ Rev. }{\bf D51}, 379(1995).

\bibitem{affleck-dine}
I. Affleck and M. Dine, {\sl Nucl.\ Phys.\ }{\bf B249}, 361(1985).

\bibitem{susy-unification}
P. Langacker and M. Luo, {\sl Phys.\ Rev. }{\bf D44}, 817(1991);

U. Amaldi, W.de Boer, and H. F\"{u}rstenau,
{\sl Phys.\ Lett.\ }{\bf B260}, 447(1991).

\bibitem{fyy81} 
M. Fukugita, T. Yanagida and M. Yoshimura, {\sl Phys.\ Lett.\ }{\bf B106},
183(1981).

\bibitem{harvey-turner}
J.A. Harvey and M.S. Turner,{\sl Phys.\ Rev. }{\bf D42}, 3344(1990).

\bibitem{gravitino problem}
J. Ellis, J.E. Kim, and D.V. Nanopoulos, {\sl Phys.\ Lett.\ }{\bf B145}, 
181(1984). \\
For recent works,
M. Kawasaki and T. Moroi, {\sl Prog.\ Theor.\ Phys.\ }{\bf 93}, 879(1995)
and references therein.

\bibitem{reheat-original}
A.D. Dolgov and A.D. Linde, {\sl Phys.\ Lett.\ }{\bf B116}, 329(1982);
L.F. Abbott, E. Fahri, and M. Wise, {\sl Phys.\ Lett.\ }{\bf B117},
29(1982).

\bibitem{landau-lifschitz m}
For a review of physical aspects, 
L. Landau and E. Lifschitz, {\sl Mechanics}
(Pergamon, Oxford, 1960), p80.

\bibitem{coddington}
For a review of mathematical aspects, 
E.A. Coddington and N. Levinson, 
{\em Theory of Ordinary Differential Equations},
(MacGraw-Hill, New York, 1955).


\bibitem{mine95-1} 
M. Yoshimura, {\sl Prog.\ Theor.\ Phys.\ }{\bf 94}, 873(1995).

\bibitem{reheating-rev-mine}
For a review,
M. Yoshimura, {\em Particle Production and Inflaton Decay},
hep-ph/9602268 and 
in {\em Cosmological Constant and the Evolution
of the Universe}, ed. by K. Sato, T. Suginohara, and
N. Sugiyama (Universal Academy Press, Tokyo, Japan, 1996).


\bibitem{reheating parametric}
A.D. Dolgov and D.P. Kirilova, {\sl Sov.\ J.\ Nucl.\ Phys.\ }{\bf 51},
172(1990);
J.H. Traschen and R.H. Brandenberger, {\sl Phys.\ Rev. }{\bf D42},
2491(1990).


\bibitem{linde et al 94}
L. Kofman, A. Linde, and A.A. Starobinsky, {\sl Phys.\ Rev.\ Lett.\ }{\bf 73},
3195(1994); hep-th/9510119.

\bibitem{holman 95}
D. Boyanovsky, M. D'Attanasio, H.J. de Vega, R. Holman, D.-S. Lee, and
A. Singh, preprint PITT-09-95; 
{\sl Phys.\ Rev. }{\bf D51}, 4419(1995) and references therein.

\bibitem{brandenberger et al}
Y. Shtanov, J. Traschen, and R. Brandenberger, {\sl Phys.\ Rev. }{\bf D51},
5438(1995).

\bibitem{linde83} 
A.D. Linde, {\sl Phys.\ Lett.\ }{\bf B129}, 177(1983).


\bibitem{mathieu eq}
A. Erdelyi, et. al., {\sl Higher Transcendental Functions} Vol.\lromn 3,
(McGraw-Hill, New York, 1955);

N.W. McLachlan, {\sl Theory and Application of Mathieu Functions},
(Oxford University Press, London, 1947).

\bibitem{fkyy95-1}
H. Fujisaki, K. Kumekawa, M. Yamaguchi, and M. Yoshimura,
{\em Particle Production and Dissipative Cosmic Field},
hep-ph/9508378; {\sl Phys.\ Rev. }, in press.

\bibitem{mine96-1}
M. Yoshimura, {\em Decay Rate of Coherent Field Oscillation},
hep-ph/9603356 and to appear in the Proceedings of the Symposium on
{\em Frontiers in Quantum Field Theory}, (World Scientific,
Singapore, 1996).


\bibitem{fkyy95-2}
H. Fujisaki, K. Kumekawa, M. Yamaguchi, and M. Yoshimura,
{\em Particle Production and Gravitino Abundance after Inflation
},
TU/95/493 and hep-ph/9511381.


\bibitem{stable gravitino}
T. Moroi, H. Murayama, and M. Yamaguchi,
{\sl Phys.\ Lett.\ }{\bf B303}, 289(1993) and references therein.


\end{thebibliography}
\end{document}